\documentstyle[12pt]{article}
\newcommand{\beq}{\begin{equation}}
\newcommand{\eeq}{\end{equation}}
\newcommand{\bea}{\begin{eqnarray}}
\newcommand{\eea}{\end{eqnarray}}

\begin{document}
\setcounter{page}{0}
\topmargin 0pt
\oddsidemargin 5mm
\renewcommand{\thefootnote}{\fnsymbol{footnote}}
\newpage
\setcounter{page}{0}
\begin{titlepage}

\begin{flushright}
QMW-PH-95-9\\
UMDEPP 95-130\\
{\bf hep-th/yymmnn}\\
May $5th$, $1995$
\end{flushright}
\vspace{0.5cm}
\begin{center}
{\Large {\bf On continuous conformal deformation of the $SL(2)_4/U(1)$
coset}} \\
\vspace{1.8cm}
\vspace{0.5cm}
{\large
S. James Gates Jr.$~^1$\footnote{e-mail: Gates@umdhep.umd.edu} and Oleg A.
Soloviev$~^{1,2}$
\footnote{e-mail: O.A.Soloviev@QMW.AC.UK}}\\
\vspace{0.5cm}
$~^1${\em Physics Department, University of Maryland, \\
College Park, MD 20742, U.S.A.}\\
\vspace{0.5cm}
$~^2${\em Physics Department, Queen Mary and Westfield College, \\
Mile End Road, London E1 4NS, United Kingdom}\\
\vspace{0.5cm}
\renewcommand{\thefootnote}{\arabic{footnote}}
\setcounter{footnote}{0}
\begin{abstract}
{We describe a one-parameter family of $c=1$ CFT's as a continuous
conformal deformation of the $SL(2)_4/U(1)$ coset.}
\end{abstract}
\vspace{0.5cm}
\centerline{May 1995}
 \end{center}
\end{titlepage}
\newpage
\section{Introduction}

The curious fact is that among all $c=1$ CFT's there exists the one-parameter
family of theories living on an orbifold \cite{Ginsparg}. These conformal
models are related to the Ashkin-Teller critical line \cite{Ashkin} and
correspondingly to the Abelian massless Thirring model. By now
many of the aspects of this $c=1$ family have been studied in detail
\cite{Ginsparg},\cite{Dijkraaf}. A new puzzle arose, when it was shown that
there is another one-parameter family of $c=1$ CFT's associated with the
affine-Virasoro construction of the affine algebra $SL(2)$ at level $k=4$
\cite{Morozov1}. According to our observation this family has to be described
as a non-Abelian Thirring model \cite{Soloviev1}. It is not obvious that the
non-Abelian Thirring model is equivalent to the Abelian Thirring model.
Therefore, one can raise a question about equivalence of the two given
one-parameter families. As yet this question has not been answered, though some
evidence for this equivalence has been presented \cite{Morozov2}.

We believe that in order to clarify the relation between the two families, one
has to compare their partition functions on the torus. The problem is that
there still are some open questions about the Lagrangian formulation of the
second continuous $c=1$ family. The aim of the present paper is to describe the
non-Abelian one-parameter family as a continuous conformal deformation of the
CFT, whose partition function on the torus is known. We shall exhibit that the
$SL(2)_k/U(1)$ coset, which is the $c=1$ CFT at $k=4$, admits a one-parameter
conformal deformation. We shall argue that the given continuous family
of $c=1$ CFT's share all the properties of the $c=1$ one-parameter
affine-Virasoro construction and is in agreement with the nonperturbative
Lagrangian description in \cite{Soloviev1}. We hope that the model we
construct will turn out to be useful for answering the main question of
equivalence.

The paper is organized as follows. In section 2 we show that the spectrum of
the $SL(2)_k/U(1)$ coset at $k=4$ acquires an extra null-vector. In section 3
we make use of the given null-vector to perform the conformal one-parameter
deformation on the $SL(2)_4/U(1)$ coset. We conclude in section 4.

\section{The null-operator at $k=4$}

The key point of our discussion will be the fact that at $k=4$ the
$SL(2)_k/U(1)$ coset acquires a new null-operator which can be used to deform
the given CFT. Therefore, we start with the Lagrangian description of the coset
construction which will be a central element in our consideration.

It is well known that within the Lagrangian approach the $G/H$ coset can be
presented as a combination of ordinary conformal Wess-Zumino-Novikov-Witten
(WZNW) models and ghost-like action \cite{Karabali}, \cite{Hwang}:
\begin{equation}
S_{G/H}=S_{WZNW}(g,k)~+~S_{WZNW}(h,-k-2c_V(H))~+~S_{Gh}(b,c,\bar b,\bar c),
\end{equation}
where $g$ takes values on the group $G$, $h$ takes values on the group $H$,
$c_V(H)$ is defined according to
\begin{equation}
f^{il}_kf^{jk}_l=-c_V(H)\delta^{ij},~~~~~i,j,k=1,2,...,\dim H.\end{equation}
Whereas the last term in eq. (2.1) is the contribution from the ghost-like
fields,
\begin{equation}
S_{Gh}=\mbox{Tr}\int d^2z(b\bar\partial c+\bar b\partial\bar c).\end{equation}
There are first class constraints in the system. Therefore, the physical states
are defined as cohomology classes of the nilpotent BRST operator $Q$
\cite{Karabali},\cite{Hwang},
\begin{equation}
Q=\oint{dz\over2\pi i}\left[:c_i(\tilde
J^i+J_H^i):(z)~-~(1/2)f^{ij}_k:c_ic_jb^k:(z)\right],~~~~~Q^2=0,\end{equation}
where we have used the following notations
\begin{equation}
J_H=-{k\over2}g^{-1}\partial g|_H,~~~~~~\tilde
J={(k+2c_V(H))\over2}h^{-1}\partial h.\end{equation}
Here the current $J_H$ is a projection of the ${\cal G}$-valued current $J$ on
the subalgebra ${\cal H}$ of ${\cal G}$.

Let us turn to the case of the $SL(2)_k/U(1)$ coset. This is the simplest coset
construction. In particular, the BRST operator $Q$ takes the form
\begin{equation}
Q=\oint{dz\over2\pi i}:c(\tilde J^3+J^3):(z),\end{equation}
where we assume that the gauge subgroup $H=U(1)$ is associated with the
subalgebra generated by the $t^3$ generator of the $SL(2)$ algebra
\begin{equation}
[t^a,t^b]=f^{ab}_ct^c,~~~~a,b,c=1,2,3\end{equation}
thus, $c_V(U(1))=0$. Therefore, the action of the $SL(2)_k/U(1)$ coset is given
by
\begin{equation}
S_{SL(2)/U(1)}=S_{WZNW}(g,k)~+~S_{WZNW}(h,-k)~+~\int d^2z(b\bar\partial c+\bar
b\partial\bar c).\end{equation}

The physical operators and states are constructed in terms of the three given
CFT's under the condition of annihilation by $Q$. We are not interested in
finding all the physical operators. We want to focus attention on the following
one
\begin{equation}
O^L=L_{ab}:J^a\bar J^{\bar a}\phi^{b\bar a}:,\end{equation}
where $\phi^{a\bar a}$ is defined as follows
\begin{equation}
\phi^{a\bar a}=\mbox{Tr}:g^{-1}t^agt^{\bar a}:.\end{equation}
This is a highest weight vector of the affine algebra \cite{Knizhnik}. In
addition, $\phi^{a\bar a}$ is a primary operator of the Virasoro algebra
associated with the WZNW model on $SL(2)_k$. The corresponding conformal
dimensions are given by \cite{Knizhnik}
\begin{equation}
\Delta_\phi=\bar\Delta_\phi={2\over k+2}.\end{equation}
In virtue of the properties of $\phi^{a\bar a}$, the operator $O^L$ is also a
Virasoro highest weight vector, when the matrix $L_{ab}$ is symmetrical, i.e.
\begin{equation}
L_0|O^L\rangle=\Delta_O|O^L\rangle,~~~~~~L_{m>0}|O^L\rangle=0,\end{equation}
with
\begin{equation}
\Delta_O=\bar\Delta_O=1+{2\over k+2}.\end{equation}
Here $L_n$ are the Virasoro generators of the WZNW model on $SL(2)_k$. At the
same time, the operator $O^L$ is no longer a highest weight vector with respect
to the affine algebra but its descendant.

In the case of $SL(2)$ the normal form of the matrix $L_{ab}$ is diagonal,
\begin{equation}
L_{ab}=\lambda_a\eta_{ab},\end{equation}
where $\lambda_1,~\lambda_2,~\lambda_3$ are arbitrary numbers,
$\eta_{ab}=\mbox{diag}(1,1,-1)$.

One can check that in general the operator $O^L$ does not commute with $Q$ and,
therefore, does not belong to the physical subspace of the theory in eq. (2.8).
Hence, some modifications of $O^L$ are required.

Let us consider the following modified operator
\begin{equation}
\hat O^L=O^L~+~N:\tilde J^3\bar
J^{\bar a}\phi^{3\bar a}:,
\end{equation}
where the constant $N$ is to be defined.
Clearly the operator $\hat O^L$ has
the same conformal dimensions as $O^L$. We demand
\begin{equation}
Q|\hat O^L\rangle=0.\end{equation}
This requirement leads us to two equations
\begin{equation}
\lambda_1=\lambda_2,~~~~~~~N=\lambda_3-{2\over
k}(\lambda_1+\lambda).\end{equation}
Under the given conditions, the operator $\hat O^L$ belongs to the cohomology
of the BRST operator Q. In general, $\hat O^L$ is built up of the two WZNW
models presented in eq. (2.8). Dramatic simplifications occur, when $k=4$.
Indeed, in this case there is a solution with
$\lambda_1=\lambda_2=\lambda_3=\lambda$ and $N=0$. Thus, at $k=4$,
\begin{equation}
\hat O^L=O^\lambda,\end{equation}
where
\begin{equation}
L_{ab}=\lambda\eta_{ab}.\end{equation}

The operator $O^\lambda$ describes a one-parameter family of BRST invariant
operators. It turns out that $O^\lambda$ is a null-vector. Indeed, it is not
difficult to compute its norm with arbitrary $k$:
\begin{equation}
||O^\lambda||^2=(k-4)^2~M,\end{equation}
where $M$ is some number. Hence, the norm vanishes, when $k=4$. Also one can
show that
\begin{equation}
\langle O^\lambda O^\lambda O^\lambda\rangle=0,\end{equation}
which agrees with the consistency equation obtained in \cite{Soloviev2}.
Therefore, all correlation functions with $O^\lambda$ will vanish. In other
words, at $k=4$
the space of physical operators of the $SL(2)_k/U(1)$ coset acquires a
one-parameter family of null-vectors which are not ruled out by the BRST
symmetry and are not equal to BRST exact operators.

\section{The one-parameter deformation}

Let us consider the following theory
\begin{equation}
S(\lambda)=S_{SL(2)_4/U(1)}~-~\int d^2zO^\lambda(z,\bar z),\end{equation}
where $O^\lambda(z,\bar z)$ is the null-operator constructed in the previous
section. The $\lambda$ parameter in eq. (3.22) has dimension $-2\Delta_\phi$,
because the operator $O^\lambda$ has dimensions
$(1+\Delta_\phi,1+\Delta_\phi)$. Since $\Delta_\phi>0$, $O^\lambda$ is an
irrelevant operator. If we think of $\lambda$ as being small, then the action
in eq. (3.22) can be understood as the $SL(2)_4/U(1)$ coset perturbed by the
irrelevant operator. In general, such a perturbation will run us in the
infrared
problem. However, in the case under consideration $O^\lambda$ is a null-vector.
Therefore, there should be no trace of this operator in any local physical
observable. Correspondingly the conformal symmetry of the perturbed theory
remains to be manifest. Thus, $\lambda$ appears to play a role of a modular
variable from the
point of view of the target space geometry described by the given CFT. This
analogy can be made more clear in terms of interacting WZNW models \cite{Hull}.

Let us consider the system of two interacting identical $SL(2)_4$ WZNW models.
The action of the system is given by
\begin{equation}
S(g_1,g_2,S)=S_{WZNW}(g_1,4)+S_{WZNW}(g_2,4)-{16\over\pi}\int
d^2z\mbox{Tr}^2(g_1^{-1}\partial g_1S\bar\partial g_2g_2^{-1}).\end{equation}
Here $g_1,~g_2$ take values in $SL(2)$. The statement is that \cite{Hull}
\begin{equation}
Z(4,4,S)=Z_{WZNW}(4)\tilde Z(\lambda),\end{equation}
where
\begin{eqnarray}
Z(4,4,S)&=&\int{\cal D}g_1{\cal D}g_2\exp[-S(g_1,g_2,S)],\nonumber\\
Z_{WZNW}(4)&=&\int{\cal D}g\exp[-S_{WZNW}(g,4)],\\
\tilde Z(\lambda)&=&\int{\cal D}g\exp[-(S_{WZNW}(g,4)-\int d^2zO^\lambda)].
\nonumber\end{eqnarray}
The relation between the couplings $S$ and $\lambda$ is as follows
\begin{eqnarray}
S&=&\sigma\cdot I,\nonumber\\ & & \\
\lambda&=&-{16\sigma^2\over\pi}+...,\nonumber\end{eqnarray}
where $I$ is the identity from the direct product of two Lie algebras, ${\cal
G}\times{\cal G}$, $\sigma$ is a small parameter, whereas
dots stand for higher order corrections in $\sigma$. The last formula
tells us that we have to restrict $\lambda$ to negative values. Furthermore, at
$\sigma=1/8$, the system of two interacting level 4 WZNW models acquires the
gauge symmetry and, correspondingly, undergoes phase transition. There is one
more phase transition at the Dashen-Frishman point \cite{Dashen} (see also
\cite{Hull}). Therefore, $\sigma$ is not exactly a continuous parameter, but
has two particular values at which the system of two interacting WZNW models
changes its properties drastically.

Now it becomes transparent how $\lambda$ can be related to the module of the
target space geometry. Indeed, the interaction term in eq. (3.23) is to
parametrize the metric on the intersection of the two group manifolds. Since we
have shown that the coupling $S$ may change continuously without changing the
underlying CFT, this parameter $\sigma$ is by definition called a module. This
fact must imply that the current-current interaction in eq. (3.23) becomes a
truly marginal operator at $k=4$.

The Virasoro central charge of the CFT described by eq. (3.23) is equal to 4.
Unfortunately,
it is not clear how the $c=1$ CFT is embedded into the given $c=4$
CFT. Based on formula (3.24) we may argue that the stress-energy tensor of the
$SL(2)_4$ WZNW model perturbed by the null-vector $O^\lambda$ has to have the
following form
\begin{equation}
T(\lambda)= T_{c=1}(\lambda)~+~K,\end{equation}
where $K$ is the $c=1$ CFT which is gauged away by gauging the subgroup
$H=U(1)$. The hope is that there exist variables in which the stress-tensor
$T_{c=1}(\lambda)$ can be presented as the one-parameter affine-Virasoro
construction \cite{Morozov1}. The latter also has two points which might be
identified with the two phase transitions we have just mentioned above.
We leave this issue for further investigation.

The fact that the system of two interacting level $k=4$ WZNW models is
conformal with the continuous parameter $\sigma$ indicates that the interaction
term in eq. (3.23) is a truly marginal operator. For all other $k$'s this is
just a marginal operator which breaks the conformal symmetry. Truly marginal
operators are responsible for the existence of moduli in the target space
geometry. This observation supports our interpretation of $\sigma$ (or
$\lambda$) as a module.

\section{Conclusion}

We have found a continuous family of $c=1$ CFT's, which either can be
associated with the truly marginal perturbation of the system of two $SL(2)_4$
WZNW models or with the null-deformation of the $SL(2)_4$ WZNW model. Both
descriptions are equivalent to each other. If the given one-parameter family is
linked with the $c=1$ orbifold construction, then one may expect a duality
symmetry to present in the theory we have constructed. Indeed, via bosonization
procedure one can relate the parameter $\sigma$ in eqs. (3.26) with the radius
$R$ of a scalar field compactified on a circle:
\begin{equation}
R=\sqrt{{(1-8\sigma)\over2}}.\end{equation}
It is well known that this compactification possesses the duality symmetry
under $R\to1/(2R)$.

Our consideration of the continuous deformation of the CFT
suggests some general arguments about perturbation theory in duality invariant
systems. Suppose $R$ is a parameter which goes to $1/(2R)$ under the duality
symmetry. (In the non-Abelian case one has some matrices instead of one
parameter.) Then there is a self-dual point $R_0=1/\sqrt2$. Near by this point
there is a small parameter
\begin{equation}
r={R-R_0\over R_0},\end{equation}
which measures the deviation from $R_0$. Under the duality symmetry
\begin{equation}
r\to r'=-r~+~{\cal O}(r^2).\end{equation}
If $r$ is very tiny, one can drop the higher order corrections in $r$ in the
equation for $r'$. Correspondingly, the original duality symmetry amounts to
the change of sign of the parameter $r$. Therefore, the leading order in
$r$ has to be an even function of $r$. In other words, expansion in $r$ has to
have an effective perturbation parameter which is an even function of $r$.

We have exhibited how quantum
perturbation theory can be related to expansion around the self-dual point,
which in the case under consideration coincides with the $SL(2)_4/U(1)$ coset.
Indeed, eqs. (3.26) clearly display the effect we anticipate for duality
invariant systems.
The hope is that in the so-called $S$-dual theories, expansion around the
self-dual point can be connected with string quantum expansion in topology of
the world sheet. This would be certainly the case, when the tree approximation
in string theory coincides with the $S$-self-dual point. Then one could check
whether the string toroidal correction gives rise to the even function of $r$.

{\bf Acknowledgements:}

S.J.G.'s research is supported by NSF grant \# PHY-93-41926 and in part by NATO
Grant CRG-93-0789.
O.S. would like to thank the British PPARC and the Physics Department of the
University of Maryland for financial support.

\end{document}